# Formation of spiral and elliptical galaxies in a CDM cosmogony

Matthias Steinmetz[1,2] and Ewald Müller[1]

[1]Max–Planck–Institut für Astrophysik,
Karl–Schwarzschild–Straße 1,
85740 Garching b. München, FRG

[2]Institut für Theoretische Physik
und Sternwarte der Universität Kiel,
Olshausenstraße 40,
24098 Kiel, FRG

Send offprint requests to: M. Steinmetz

Thesaurus codes: 03.11.1, 07.08.1, 07.14.1, 07.22.1, 07.23.1

Submitted to: *Astronomy and Astrophysics Letters*

## Abstract

The formation of galaxies in a *Cold Dark Matter* cosmogony is investigated by following the evolution of dark and baryonic matter and of the frequency-dependent spatially averaged radiation field. The gas is allowed to form stars which are independently treated as a fourth collisionless component. By supernovae, energy and metal enriched gas is returned to the interstellar medium. Two extreme cases, the formation of an isolated field galaxy and the merging of two gas rich proto spirals are considered the latter being used as a model for galaxy formation in a cluster.

The isolated galaxy is assumed to interact with the surrounding matter via tidal fields only, resulting in a rotation with a spin parameter $\lambda \approx 0.08$. The forming field galaxy shows the main properties of spiral galaxies: A metal rich bulge, a metal poor stellar halo and a disk of nearly solar composition. The spiral is embedded in a triaxial halo of dark matter. The disk possesses an exponential brightness profile, while halo and bulge follow the de Vaucouleurs law. The disk is rotationally supported with a flat rotation curve. The flattening of the triaxial bulge is probably due to its rotation, whereas the triaxial halo rotates only slowly. The disk has a metallicity gradient of $d(\log Z)/dr = 0.05/\mathrm{kpc}$, whereas the halo shows none. The models also exhibit a correlation between the disk to bulge ratio and the power of small scale fluctuations. The stars of the bulge form from gas which is initially located in the largest maxima of the primordial density fluctuations, whereas the halo stars originate from gas accumulated in less pronounced maxima.

Merging proto spirals evolve to objects which can be interpreted as luminous ellipticals possessing a de Vaucouleurs profile. The metal rich ($Z \approx 2\,Z_\odot$) elliptical rotates only slowly its flattening being a result of an anisotropic velocity dispersion. The metallicity of the elliptical decreases with radius according to a power law. Depending on the geometry of the merger event sometimes counter rotating gas is observed.

## 1. Introduction

The explanation of the origin of the observed different morphologies of galaxies poses one of the greatest challenges in the field of cosmology and galactic dynamics. In particular, the origin of the two major types of galaxies, ellipticals and spirals, has been and still is a matter of controversial debates. From observations one knows that spiral galaxies are rotationally supported and that their surface brightness follows an exponential law. Their rotation curves which are constant out to radii of 50 kpc or even beyond cannot be explained by galactic models which only consider visible matter. Observations further show that many spirals possess a more or less pronounced central spheroidal bulge, which like elliptical galaxies, is stabilized by its velocity dispersion, and that the surface brightness of elliptical galaxies can well be described by a de Vaucouleurs profile ($\log I \propto r^{(1/4)}$). Observations by Bender (1989) imply that there exist two distinct classes of luminous elliptical galaxies having *boxy* and *disky* isophotes. In general, boxy ellipticals seem to be more massive ($M_B < -20.5$) and being flattened by an anisotropic velocity dispersion. In contrast, the ellipticity of generally less massive disky ellipticals ($M_B > -20.5$) can easily be explained by their rotation. Furthermore, it is interesting that the bulges of spiral galaxies, which also show a de Vaucouleurs profile and have masses comparable to disky ellipticals, seem to be rotationally flattened, too.

During the last two decades several models have been proposed to explain the morphological differences between ellipticals and spirals. However, most of those models ignored the influence of the underlying cosmogony for the structure formation. This drawback was the focus of the pioneering work of Katz (1992), who simulated the formation of spirals in a hierarchical cosmogony including gas dynamics and star formation. In his simulations objects formed which have the main properties of observed spiral galaxies: a disk, a stellar spheroid and a dark matter halo. On the other hand computer simulations of merging spiral galaxies (see Barnes & Hernquist 1992) show the formation of an object which resembles the observed structure of high mass ellipticals except for their compactness.

In this letter we present the first results of a comprehensive study which extends the work of Katz (1992) as well as that of Barnes & Hernquist (1992). We have simulated the evolution of isolated spiral galaxies of different mass, angular momentum and star formation history resulting from different primordial fluctuation spectra. Besides providing a more detailed analysis of the structure and the kinematics of the forming galaxy, the simulations also incorporate the metal enrichment by supernovae in a simple form, which allows us to identify different stellar populations. In order to study galaxy formation in a cluster environment we have also simulated the merging of two gas rich star forming proto spirals instead of two fully evolved spirals as in the work of Barnes & Hernquist (1992).

## 2. Method

The simulations have been performed with a combined smoothed–particle–hydrodynamics (SPH) and N–body tree code (Steinmetz & Müller 1993). Radiative cooling is described by a time-dependent ionization network including the ions of H and He. The evolution of the spatially averaged frequency-dependent radiation field is treated as in Cen (1992), and star formation is included in a way similar to that described by Katz (1992). Once stars are formed their further evolution is followed by a N–body code. It is assumed that they are distributed according to a Miller–Scalo (1979) initial mass function. Stars more massive than $8\,M_\odot$ explode as Type II supernovae with an energy release of $10^{51}$ erg and return, except for an $1.4 M_\odot$ compact remnant, all their mass to the gaseous component. It is further assumed that 15% of the ejected mass are metals.

In a typical simulation 4000 dark matter and 4000 gas particles are used. During a simulation typically 25 000 to 35 000 stellar particles are formed. For numerical reasons, the (adaptive) softening length of the gravitational potential is restricted to be larger than 1 kpc (stars), 1.5 kpc (gas) and 3 kpc (dark matter), respectively. A typical run requires between 50 h (spirals) and 150 h (ellipticals) CPU time on one processor of a CRAY YMP. For a more detailed discussion of the method and the results we refer to forthcoming publications (Steinmetz & Müller 1993a,b,c in preparation).

## 3. Spiral galaxies

All calculations start at a redshift of $z = 25$, when baryons and dark matter are still equally distributed. The initial model consists of a homogeneous sphere with a radius of 50 kpc on which small scale fluctuations are superimposed according to the Zel'dovich approximation. This approach neglects all power on scales larger than the box volume. However, the analytical properties of the gravitational clustering process allow one to derive some approximations for the influence of the large scale power. As long as no significant amount of matter is transported into the simulated cosmic subvolume, the large scale power can be split into two main effects: a compression of the volume and a tidal field which causes the proto galaxy to rotate. Both the amount of compression and rotation can be predicted via perturbation theory. For a CDM scenario a density enhancement of a factor of about 1.3 compared to the mean cosmic density is expected for $z = 25$. The rotation of the proto galaxy is usually measured by the dimensionless spin parameter $\lambda = J\sqrt{|E|}/(G\,M^{2.5})$, where $J$, $E$, $G$ and $M$ are the angular momentum, the total energy, the gravitational constant and the mass, respectively. Relatively independent on the special form of the primordial fluctuation spectrum a spin parameter of $\lambda \approx 0.08$ can be obtained (Barnes & Efstathiou 1987). Therefore, realistic initial conditions for the formation of a field galaxy are a rotating, density enhanced sphere which expands according to the Hubble law. However, the expansion velocity of the sphere should be reduced according to the density enhancement. The expansion will end at $z \approx 5$ followed by a collapse phase. As in the work of Katz (1992), simulations with these initial conditions lead to the formation of a spiral galaxy.

The evolution can be separated into several phases. For $z \gtrsim 10$ the evolution of the dark and baryonic matter is identical. Later, the gas cools and forms clumps which are denser than the dark matter. Near the maxima of the primordial density fluctuations the gas begins to form stars. Up to a redshift of $z \approx 2.5$ the most pronounced maxima merge and form an object of ellipsoidal shape which is surrounded by more diffusely distributed stars. The star formation rate during this phase is of the order of $20\,\mathrm{M}_\odot$/yr. Between a redshift of 2.5 and 1.8 the disk begins to form by an accretion like process and the star formation rate peaks at about $50\mathrm{M}_\odot$/yr. At $z \approx 1.8$ the star formation rate begins to drop, the disk component becomes more and more dominant and at $z = 1$, when the calculation was stopped, a quasi–stationary state is reached. The mass fraction of the gas in the disk has dropped to 20 to 30% and the star formation rate has leveled of at about $1\mathrm{M}_\odot$/yr.

Dividing the stellar component of the final object into two subsamples, old stars ($t_* > 2.5\,\mathrm{Gyr}$ at $z = 1$, i.e., $t_* > 11\,\mathrm{Gyr}$ at $z = 0$) and new stars ($t_* < 0.5\,\mathrm{Gyr}$), the calculations show that the stars of the galaxy form three distinct spatial components, the disk, the bulge and the stellar halo. The young stars are preferentially located in a thin stellar disk whose scale height increases with the age of the stars. The surface density of the stellar disk follows an exponential law with a scale length of 3.5 kpc. The scale length of the gaseous disk is $\approx 6$ kpc. The disk is rotationally supported and the stars rotate with about 220 km/sec, which is in agreement with the circular velocity of the underlying mass distribution. The rotation curve is flat up to a distance of 30 kpc. The velocity dispersion of the disk stars is rather small, i.e., the disk is cold, and tends to increase with the age of the stars. Typical values are of the order of 20 km/sec. The disk stars have a metallicity $0.1Z_\odot < Z < 2Z_\odot$, which decreases exponentially with a gradient $d\log Z/dr = -0.05$/kpc (Fig. 1) in good agreement with observations.

The properties of the old stars are rather different. They form the bulge–halo system, which is of ellipsoidal shape and exhibits a de Vaucouleurs profile. The bulge is composed of old, metal rich stars ($Z \approx 2\,Z_\odot$), whereas the halo consists of old, metal poor stars ($Z < 0.3\,Z_\odot$). The halo component shows nearly no metallicity gradient for $5\,\mathrm{kpc} \lesssim r \lesssim 20\,\mathrm{kpc}$ (Fig. 1), which is in agreement with observations of Zinn (1985). Compared with the rotation velocity of the disk, the rotation velocity of the bulge–halo system is rather small. Starting in the center the rotation curve rapidly rises to a value of about 100 km/sec at a radius of about 1 kpc. Further out the rotation curve decreases, and becomes nearly constant beyond a radius of 10 kpc at values of about 50 to 70 km/sec. On the other hand the velocity dispersion of the bulge–halo system is larger than that of the disk. Typical values are 120 km/sec for the radial and 70 km/sec for the tangential velocity dispersion the anisotropy increasing with radius. Therefore, the bulge–halo system is pressure supported. The flattening of the halo (axial ratio $\approx$ 2:3) is too high to be a result of its rotation, i.e., it must be caused by the anisotropic velocity dispersion, while the flattening of the bulge (axial ratio 1:1.3) is at least partially due to its rotation ($v/\sigma \approx 1$). These results support the idea that *disky* ellipticals and the bulges of spiral galaxies have the same origin.

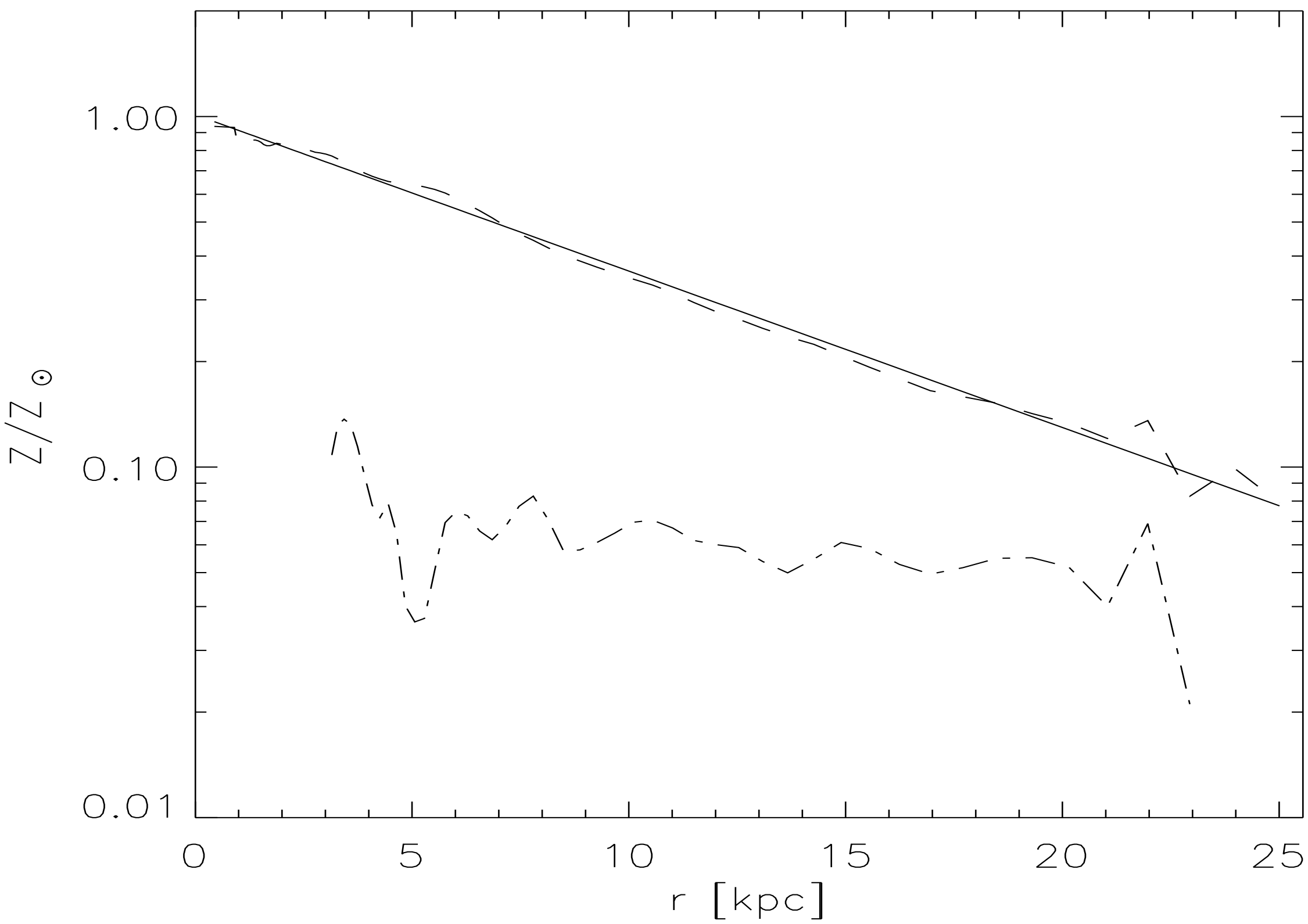

Figure 1: Metallicity of the stars as a function of radius. The dashed line shows the metallicity of the disk stars, the dash–dotted that of the halo stars ($|z| > 3\,\mathrm{kpc}$, $t_* > 2\,\mathrm{Gyr}$). The solid line represents a least square fit of the metallicity of the disk stars the gradient being $d(\log Z)/dr = -0.05/\mathrm{kpc}$. Note that the halo stars posses nearly no metallicity gradient.

The three stellar components also show up in the star formation history, which can be divided into two phases. In the first phase lasting from 4 to 2 Gyr before $z = 1$, the bulge and halo stars are formed. Within only 500 Million years the metallicity near the center rapidly increases towards $Z = 2\,Z_\odot$, while the metallicity of the more diffusively distributed halo stars remains rather small. The left frame of Fig. 2 shows the distribution of the gas particles at a redshift of $z \approx 6$. If instead only those gas particles are plotted, which will form old but metal rich stars (i.e., bulge stars), one finds that they are located in the inner 1 to 2 kpc of the final object (middle frame of Fig. 2). Thus, bulge stars are preferentially formed in the main maximum of the primordial fluctuation field. On the other hand gas particles, which end up in old and metal poor stars, are more diffusively distributed (right frame of Fig. 2). Obviously, there exists a strong correlation between the metallicity of the older stars and the strength of the primordial density fluctuations. The same correlation is found for the metallicity of the gas component.

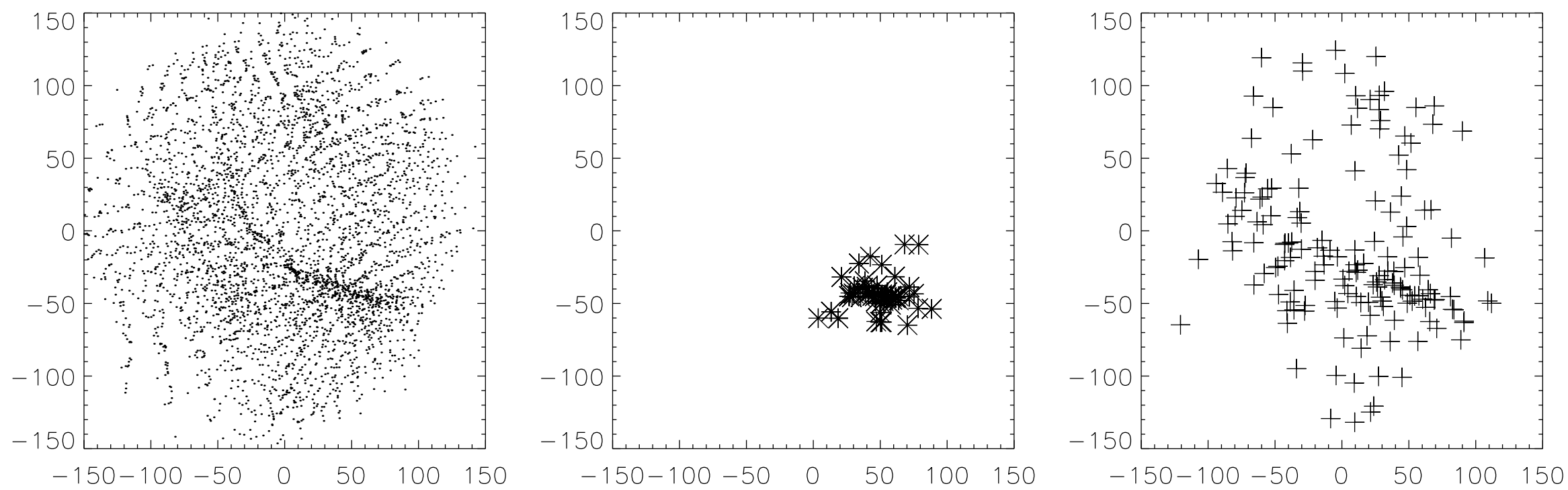

Figure 2: Snapshot of gas particle distribution in a forming spiral galaxy at a redshift of $z = 6$. Left frame: Projected distribution of all gas particles in the x–y–plane. Middle frame: Position of gas particles which will form old and metal rich (bulge) stars. Right frame: Position of gas particles which will form old and metal poor (halo) stars.

The second star formation phase begins about 2 billion years before $z = 1$, and is a main result of the collapse of the whole protogalactic cloud giving rise to the formation of the disk component. Our results show that the metallicity of the disk stars increases more slowly than that of the bulge and halo stars, and that no correlation between the metallicity of the disk stars and the primordial fluctuation power of the precessor gas particles exists.

Concerning the dependence of the results on the model parameters the primordial fluctuation spectrum seems to play the most important role. In most calculations we find a typical disk to bulge ratio D/B of 1:2 to 1:4. However, in one model, which accidentally has more small scale power, a more pronounced bulge (D/B=1) can be observed. Furthermore, in models without any small scale power, i.e., the collapse of a homogeneous rigidly rotating sphere, a disk galaxy without any bulge is formed. The old but metal rich star component is missing in these models, too. The spin parameter also influences the formation of the galaxy. It determines the scale length of the disk, which varies roughly linearly with the initial spin parameter. Even for spin parameters as low as $\lambda = 0.02$ the formation of a disk galaxy is observed. The scale length of such a disk ($\approx 1\,\mathrm{kpc}$) is, however, at the limit of the numerical resolution. The parameters determining the star formation law and the initial mass function influence the evolution only weakly, i.e., the star formation is self-regulated.

## 4. Elliptical galaxies

To simulate galaxy formation in a cluster, we put two of the rotating and collapsing spheres discussed in the last section onto a parabolic orbit varying the relative orientation of the intrinsic and orbital angular momentum. The two spheres approach the perigalacticon at redshifts between 1.8 and 2.5. The spin parameter of the whole system is of the order of $\lambda \approx 0.05$ and, thus, comparable to the spin parameter of the proto spirals discussed before! When the proto spirals begin to merge around $z = 2.5$, the disk component is not yet evolved. Consequently, a lot of energy is dissipated during the merging event giving rise to a burst like star formation which peaks at $z \approx 2.5$ with a rate of 250 up to $300\,M_{\odot}/yr$. At the end of the merging process ($z \approx 1.8$) an object is formed which shows the main properties of observed giant ellipticals: A triaxial ellipsoid with a de Vaucouleurs surface density profile. The effective radius is about 1 kpc, and thus too small. Obviously, too much energy is dissipated during the merger event. A better agreement with observation can be achieved, if the merging occurs later. In a model where the perigalacticon is reached at a redshift of 1.8, the effective radius is about 1.5 kpc.

From the kinematical point of view all mergers are slow rotators. The rotation velocity of the stars is of the order of 30 km/sec, whereas the velocity dispersion reaches values of $\approx 150$ km/sec, i.e., $v/\sigma \approx 0.2$. The velocity dispersion is anisotropic the radial velocity dispersion typically being 50% larger than the tangential velocity dispersion. Therefore, the flattening of the ellipsoid (axial ratio typically 1:1.4) is caused by the anisotropic velocity dispersion. The rest gas is rotationally supported. Depending on the relative orientation of the spin of the proto spirals and of the orbital angular momentum, counter rotating gas can be observed. However, only a small fraction of the gas ($\approx 5\%$) is located within the inner 5 kpc.

The metallicity of the stars is generally high ($Z \approx 2\,Z_{\odot}$). For radii $0.5\,R_e \lesssim R \lesssim 3\,R_e$ the metallicity of the stars can well be described by a power law $d\log Z/d\log R =$ const, the constant ranging from $-0.25$ to $-0.4$ for a merger event at $z = 1.8$ and $z \approx 2.5$, respectively. Compared with observations, the latter gradient seems to be somewhat too steep which is consistent with the too compact structure of the early merger. On the other hand the structure of the later merger is in rough agreement with observations. Note, however, that even the later merger is nearly relaxed at $z \approx 1.5$, i.e., relative to $z = 0$ the star formation ends 10 billion years ago.

## 5. Conclusion

We have performed a set of three-dimensional simulations of the formation of galaxies, which take into account the evolution of a collisionless dark component, a dissipative gas component and a stellar component formed by Jeans unstable gas. Two scenarios, the formation of an isolated spiral and the dissipative merging of two proto spirals are considered. In contrast to most previous studies, the initial conditions are consistent with the predictions of cosmological scenarios for hierarchical structure formation. In particular, small scale fluctuations with a wavelength less than the simulation volume are accurately included. The outcome of the merger event is an object which qualitatively shows the properties of observed massive ellipticals. The properties of the forming isolated spiral galaxy agree even quantitatively with those of observed spirals. Discriminated by their structure, kinematics and chemical properties the spiral shows three different components: A rotationally supported exponential disk composed of stars with solar metallicity, a metal rich moderately rotating pressure supported bulge and a slowly rotating metal poor stellar halo. The surface brightness of the halo and the bulge follows a de Vaucouleurs law. The spiral galaxy is embedded in a dark halo causing the flat rotation curve of the disk component. Thus, our results support the idea, that spiral galaxies are the standard outcome of the galaxy formation process in a CDM cosmogony, and that high mass ellipticals are the result of a merger event. The close similarity between the bulges of spirals and rotationally flattened elliptical galaxies further suggests that both originate from local maxima of the primordial density fluctuation field.

**Acknowledgements.** We would like to thank M. Bartelmann, A. Burkert, G. Hensler and C. Theis for many fruitful discussions. This work is partially supported by the *Deutsche Forschungsgemeinschaft*. All calculations were performed on the Cray YMP 4/64 of the Rechenzentrum Garching.